\documentstyle[sprocl,graphicx]{article}
\def\be{\begin{equation}}
\def\ee{\end{equation}}
\def\bea{\begin{eqnarray}}
\def\eea{\end{eqnarray}}

   \newcommand{\pr}[1]{^{prime}}

   \newcommand{\triplint}{\int\rule{-3.5mm}{0mm}\int\rule{-3.5mm}{0mm}\int}
   \newcommand{\doublint}{\int\rule{-3.5mm}{0mm}\int}

   \newcommand{\vecb}[1]{\mbox{\bf#1}}

   \newcommand{\mra}  {\rightarrow}
   \newcommand{\mlora} {\longrightarrow}

\begin{document}

\title{Microcanonical Thermodynamics, Fragmentation ``Phase-Transition''
\footnote{We use quotation marks to emphasize the limited analogy to standard
phase transitions, CRIS96,Catania May 27-31,1996.``Critical Phenomena and Collective
Observables''}, and the 
Topology of the N-body Phase Space}

\author{ D.H.E. Gross}

\address{Hahn Meitner Institut Berlin, Bereich TV,\\and Fachbereich Physik der
Freien Universit\"at\\Glienickerstr. 100, D14109 Berlin, Germany}


\maketitle\abstracts{The general features of Microcanonical Thermodynamics ({\em MT})
as applied to the fragmentation of hot nuclei and atomic clusters are
discussed.  {\em MT} is the most fundamental form of any thermodynamics since
Boltzmann. With modern computational techniques it is for the first time
possible to explore it in nontrivial cases. First and second order phase
transitions in finite systems can unambiguously be identified by the caloric
equation of state T(E/N). The three characteristics of phase transitions: The
transition temperature, the latent heat and the {\em interphase surface
tension} can be well defined and calculated in finite systems. {\em Against
common believe and in contrast to conventional (canonical) thermodynamics, {\em
MT} of relatively small systems reflects the thermodynamic behavior of bulk
systems in great detail and surprising accuracy.} By the Laplace transform from
the microcanonical to the canonical ensemble many of the signatures of phase
transitions are smeared out or get lost. Also some of our believes about phase
transitions must be corrected:  Microcanonical phase transitions show distinct
peculiarities very different from conventional ph-tr.: E.g. the specific heat
is {\em negative}. Phase separation is usually unsuitable to identify ph-tr.
in small systems and in sharp contrast to the caloric equation of state.  In
contrast to ordinary (canonical) thermodynamics {\em MT} allows the system to
become {\em inhomogeneous} or to {\em fragment}, at first order phase transitions it
gives insight into
the coexistence region.  Here the form of the specific heat $c(E/N)$ connects
transitions of first and second order in a natural way.  The ``phase transition''
towards fragmentation is introduced. The similarities and {\em differences} to
the boiling of macrosystems are pointed out.\\~\\ Nuclear friction is likely
responsible for the ergodic expansion of a multifragmented nuclear source up to
freeze-out densities of $\sim 1/6$ normal density. Various recent experimental
results which seem to contradict our scenario are even confirming it when
investigated more thoroughly.  The fragmentation ``phase-transition''
(multifragmentation) in nuclei is by many reasons {\em not} the same as the
liquid-gas transition in nuclear matter. The main reasons were already
published 1984.\cite{gross69} }
\section{Introduction}

Our definition of Microcanonical Thermodynamics ({\em MT}) uses only
mechanics. The topology of the total accessible N-body phase space reflects
(or implies?) the behavior of many interacting N-body systems because their
dynamics is often chaotic. Then the dynamical evolution of many replica of the
same system under identical macroscopic initial conditions follows the
structure of the underlying N-body phase space.  It is ergodic.  In nuclear
fragmentation this is presumably due to the strong and short ranged friction
between moving nuclei in close proximity. Friction between atomic clusters is
yet unknown but quite likely it exists there also.

First we have to discuss the concept of thermodynamics of small systems in
general and especially of phase transitions. The first question is from which
size on do phase transitions exist in small systems? How can one define them?
Is a cluster of $\sim 100$ particles big enough? We will show that this is
possible and one can unambiguously distinguish continuous (second order
transitions) from discontinuous (first order) transitions by the form of the
caloric equation of state $T_{thd}(E)$. Before proceeding further it is
important to realize that isolated nuclei or clusters must be treated {\em
microcanonically.} Usually there is no external heat- or particle bath which
defines the temperature, the pressure, or the chemical potential.
Microcanonical Thermodynamics is the proper theory for isolated small systems.

A microcanonical ensemble has some peculiarities: It does not have a positive
definite heat capacity. In fact at a phase transition of first order the
specific heat $c(\varepsilon)$ becomes {\em negative} in general. Therefore the
classical signal of a peak in the specific heat is not useful to characterize a
phase transition in small systems.

The second peculiarity of Microcanonical Thermodynamics is that it allows the
system to become {\em inhomogeneous}: At first order phase transitions several
regions of one phase coexist with other regions of the other phase. {\em MT}
allows differently to conventional thermodynamics the coexistence of regions
with different energy-density. The partitional entropy of the spatial
fluctuations is an important part of the total entropy. Small many-body systems
and also large systems under long-range forces have an important new structural
``phase transition'' which does not exist in infinite homogeneous systems: They may
{\em fragment} into few relatively {\em large} pieces. Typical examples are
nuclear multifragmentation, as was predicted very early
\cite{gross45,bondorf81,gross56}, but also the fragmentation of atomic
clusters.

In this case the size fluctuations of the fragments at the transition are of
the order of the size of the system itself. Then one cannot ignore the
``droplets'' compared to the nucleonic or monatomic vapor anymore. The ``phase
transition'' is not determined alone by the equilibrium of the homogeneous liquid
with the homogeneous gas, which in conventional (canonical) thermodynamics is
controlled by the equality of the chemical potentials of liquid and gas as we
are used to in conventional (grandcanonical thermodynamics).  Often
the number of droplets is similar or larger than the number of free nucleons or
monomers. This is one of the main lessons relevant for general physic which can
be learned from nuclear fragmentations.
\section{Microcanonical Thermodynamics}

Microcanonical Thermodynamics explores the topology of the N-body phase space
and determines how the volume $\Omega_N$ of the accessible phase space --- more
precisely the number of quantum states --- depends on the fundamental globally
conserved quantities of total energy $E=N*\varepsilon$, angular  momentum
$\vecb{L}$, mass (number of atoms) $N$, charge $Z$, linear momentum $\vecb{p}$,
and last not least the available spatial volume $V$ of the
system. This definition is the basic starting point of any thermodynamics since
Boltzmann.\cite{boltzmann}

The entropy is defined as the logarithm of $\Omega$
\begin{equation}
S(E,V,N)=Ns(\varepsilon=E/N)=ln(\Omega(E,V,N))\label{entropy}
\end{equation} and the thermodynamic temperature $T_{thd}$ is defined by
\begin{eqnarray}
\beta&=&\frac{\partial S(E,V,N)}{\partial E}= \frac{\partial s(\varepsilon)}
{\partial \varepsilon} \label{beta} ,\\
T_{thd}&=&\frac{1}{\beta} .\label{temperature}
\end{eqnarray}
By Laplace transform of $\Omega(E,V,N)$ one steps from the ``extensive''
variables like 
$E,V,N$ to the intensive ones like $T,P,\mu$. E.g. the Gibbs grand partition
function and the Gibbs grand potential are then
\begin{eqnarray} 
Z(\beta,P,\mu)&=&
\triplint_0^{\infty}{\Omega(E,V,N)e^{-\beta(E+PV-\mu N)}\;dE\;dV\;dN} ,\\ 
G(\beta,P,\mu)&=&-T\; ln[Z(\beta,P,\mu)].
\end{eqnarray}We take Boltzmann's constant $k=1$. In the same way one gets the
canonical partition function and the free energy as
\begin{eqnarray}
Z(\beta,P,N)&=&\doublint_0^{\infty}{\Omega(E,V,N)e^{-\beta N(\varepsilon+Pv)}\;dE\;dV}
,\label{laplace}\\ 
F(T,P,N)&=&-T\; ln[Z(\beta,P,N)]. 
\end{eqnarray}
\section{Differences between microcanonical and canonical ensemble}

According to van Hove  a system of $N$ particles interacting
via short range two-body  attractive forces with hard cores is
thermodynamically stable, the thermodynamic limit of $N, V
\mra\infty|_{N/V=\varrho}$ exists for such systems, intensive quantities like
the specific energy have finite limiting values.\cite{vanhove49} Then, the
thermodynamics derived from the microcanonical partition sum $\Omega(E,N,V)$
and the one derived from the canonical $Z(\beta,P,N)$ or the grand canonical
partition function $Z(\beta,P,\mu)$ (usually) coincide. Outside of phase
transitions of first order, the relative fluctuations $\Delta E/E$, or
$\Delta\varrho/\varrho$ vanish $\propto 1/\sqrt{N}$.

This is quite different {\em at} phase transitions ($T=T_{tr}$) of first order
and for {\em finite} systems where the microcanonical and the (grand)canonical
ensemble differ essentially.  In the canonical ensemble the energy fluctuations
$\Delta \varepsilon$ per particle remain finite even in the thermodynamic
limit.  ($(\Delta\varepsilon)^2|_{T_{tr}} \propto q_{lat}$, the specific latent
heat).  Consequently, the difference between the microcanonical and the
canonical ensemble persists at transitions of first order and we must expect
both ensembles describe different physical situations. 

Systems interacting via long range forces like unscreened Coulomb or the
centrifugal force when they are rapidly rotating don't have a thermodynamic
limit and must be described by the microcanonical ensemble. Such systems
fragment macroscopically into, in general several, regions of high density ---
condensed matter --- and also into, in general several, regions of low density
--- vapor or may be empty space.  Differently from conventional thermodynamics
where systems which must be in a homogeneous phase at fixed temperature
everywhere, here the system is most likely inhomogeneous.  The inhomogeneities
and their fluctuations are more important in characterizing the state of the
system than any mean values.  {\em In contrast to thermodynamics of the
homogeneous bulk, in small systems or large systems with long range forces the
entropy connected to different partitions of the system is an important part of
the total entropy.} Familiar formulas like the one-particle entropy
\begin{equation}
s_{sp}= -\sum_a{n_a ln(n_a)}
\end{equation}
are useless for calculating the total entropy.
\section{First and second order phase transitions in small systems}
 
Macroscopic systems have a discontinuity in the specific heat $c_{bulk}(T)$ at
first order phase transitions. $c_{bulk}(T)$ may have a finite peak at
$T\approx T_{tr}$.  On top of this there is a spike $q_{lat}\delta(T-T_{tr})$.
With finite resolution it shows up as jump in $c_{bulk}(T)$ by the latent heat
$q_{lat}$.  A typical example is the specific heat of bulk sodium, fig.(1) at
the melting transition.  In
contrast, a transition of second order is continuous at the transition
temperature where $c_{bulk}(T)$ has (in the example of the Ising model) a
logarithmic singularity in $T-T_{tr}$. \\
\noindent
\begin{minipage}[b]{7.5cm}
\includegraphics*[bb = 77 18 494 603, angle=-90, width=7cm,  
clip=true]{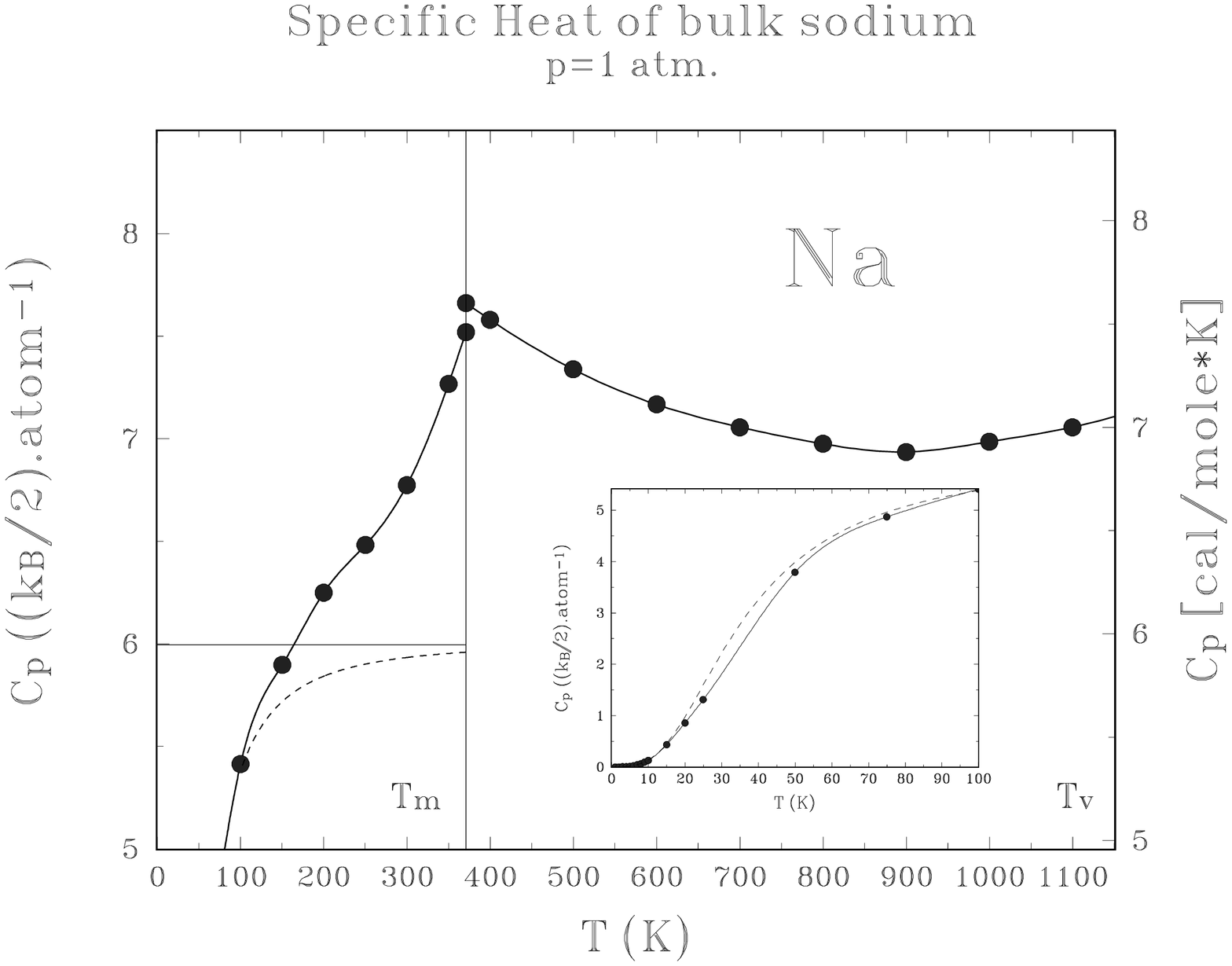}\\~\\
\scriptsize Fig.(1): Specific heat of bulk sodium at atmospheric pressure from  
\protect\cite{borelius63,hultgren63}. The dashed line represents the specific
heat calculated within the Debye model. The insert is a blow-up for 
$0\le T \le 100$K.
\end{minipage}\ ~ \
\begin{minipage}[b]{4cm}
Microcanonical~~Thermodynamics gives new and deep insight into this. It will
further allow to extend the concept of phase transitions to systems not treated
before by thermodynamics like systems with long rang forces or strongly
rotating systems. We begin with the discussion of microcanonical phase
transitions in standard model systems in which phase transitions of first and
second order are\end{minipage} well known. As example
 we take the
2-dimensional 10-states Potts model for which the asymptotic thermodynamics is
even known analytically.\cite{baxter73} 
In ref.\cite{gross150} we determined the three basic parameters of phase
transitions of first order within Microcanonical Thermodynamics, the transition
temperature $T_{tr}$, the specific latent heat $q_{lat}$, and the specific
interphase surface entropy $\Delta s_{surf}$ for a system with nearest neighbor
couplings. It was demonstrated that for surprisingly small systems the values
of these three parameters are closer to their asymptotic values than in the
canonical ensemble. This is so because most of the finite-size scaling is due
to the large, but trivial, exponent in the Laplace transform eq.(\ref{laplace})
from the micro- to the canonical partition
sum.\cite{hueller94,promberger96}

The two types of phase transitions are distinguished by the form of the
microcanonical caloric equation of state $T_{thd}(\varepsilon)$: A transition
of first order has a {\em backbending} caloric equation of state
$T^{-1}=\beta(\varepsilon)$ c.f.  figure (2b). For a system with
nearest neighbor interactions the area between $\beta(\varepsilon)$ and the
``Maxwell'' line $\beta=1/T_{tr}$ is twice the interphase surface entropy
$\Delta s_{surf}$.\cite{gross150} The left darkened area is the defect of
entropy $\Delta s_{surf}=\int_{\varepsilon_1}^{\varepsilon_2} \beta
d\varepsilon$ that the system `pays' for introducing interphase surfaces ,
which it finally gets back when the whole system is converted to the new phase
at $\varepsilon=\varepsilon_3$, right darkened area, and the interphase surface
disappeares.\cite{gross150} In the bulk the transition is discontinuous as
function of $T$ or $\beta$ and ``jumps'' from the liquid
($\varepsilon\le\varepsilon_1$) to the gas branch
($\varepsilon\ge\varepsilon_3$) of the caloric curve, fig.(2).  As a function
of the specific energy $\varepsilon$ the transition is however continuous. With
rising $\varepsilon$ the system passes smoothly from the liquid phase over a
mixed phase with coexisting large fluctuations of the two phases (``gas
bubbles'' and ``liquid droplets'') to the pure gas phase when the specific
energy is increased by the specific latent heat $q_{lat}$.  Figure (2a) shows
the corresponding specific entropy
$s(\varepsilon)=\int_0^{\varepsilon}{\beta(\varepsilon')d\varepsilon'}$.  The
transition is characterized by the convex intruder in $s(\varepsilon)$ of depth
$\Delta s_{surf}$. Figure (2c) shows the specific heat capacity
\begin{equation}
c(\varepsilon)=\frac{\partial\varepsilon}{\partial<T>}=\frac{-\beta^2}
{\partial\beta/\partial\varepsilon}
\end{equation}
as a function of the specific energy $\varepsilon$. (Here numerical
fluctuations in $\beta(\varepsilon)$ in figure (2b) have been
smoothed). One can see {\em within the coexistence region of
$\varepsilon_1\le\varepsilon\le\varepsilon_3$ (shaded area in fig
(2) ), the microcanonical specific heat has two poles and becomes
negative in between.}

As the convex intruder in the specific entropy $s(\varepsilon)$, fig.(2a), is
forbidden by van Hove's theorem in the canonical ensemble for an infinite
number of particles, conventional thermodynamics of the bulk is blind in this
energy interval.\cite{vanhove49} Here the Laplace transform eq.(\ref{laplace})
has additional stationary points and the canonical bulk would be
unstable.\cite{hueller94} It can only see the branches of 
$c(\varepsilon)$ in the regions $\varepsilon\le\varepsilon_1$ and
$\varepsilon\ge\varepsilon_3$.  Thus the canonical specific heat $c_{bulk}(T)$
will be positive and approach finite values at the transition of first order.
At $T=T_{tr}$ $c_{bulk}(T)$ has an additional peak $=q_{lat}\delta(T-T_{tr})$.

If the specific latent heat $q_{lat}\mlora 0$ and the specific interphase
surface entropy $\Delta s_{surf}\mlora 0$ the caloric equation of state gets
only a saddle point at the transition. Then $E(T)/N$ as well as
$T(E/N=\varepsilon)$ become single valued, the transition is continuous in the
canonical as well as in the microcanonical ensemble. We have a phase transition
of second order.  The two poles of $c(\varepsilon)$ merge and $c(\varepsilon)$
or $c_{bulk}(T)$ has a singularity at the transition point
$\varepsilon_{tr}$,$T_{tr}$. Both slopes of $c(\varepsilon)$ are fully
accessible in the canonical treatment of the heat capacity.
Consequently, from the caloric equation of state $T(\varepsilon)$ it is always
possible to identify and distinguish both kinds of transitions. In
Microcanonical Thermodynamics the relation between the two is very natural,
transparent and simple.

It is further instructive that in finite realizations of the two-dimensional
Potts model with $q=10$ spin orientations at each lattice point it was
not possible to see a clean separation into a compact region of ordered spins
and a compact region of disordered spins at energies inside the coexistence
region even for a lattice of $100*100$ points. There were always several ``gas
bubbles'' and ``droplets'' fluctuating over the lattice and prohibiting large
interphase surfaces. Nevertheless the caloric equation of state
$T(\varepsilon)$  is already close to its asymptotic form.  We can conclude
from this observation that the other classical signal of a transition of first
order, {\em a clear separation of the two phases}, is not a useful signal of a
transition of first order in small systems.\\~\\
\noindent
\begin{minipage}[t]{5.5cm}
\includegraphics*[bb = 79 57 375 568, angle=-180, width=5cm,  
clip=true]{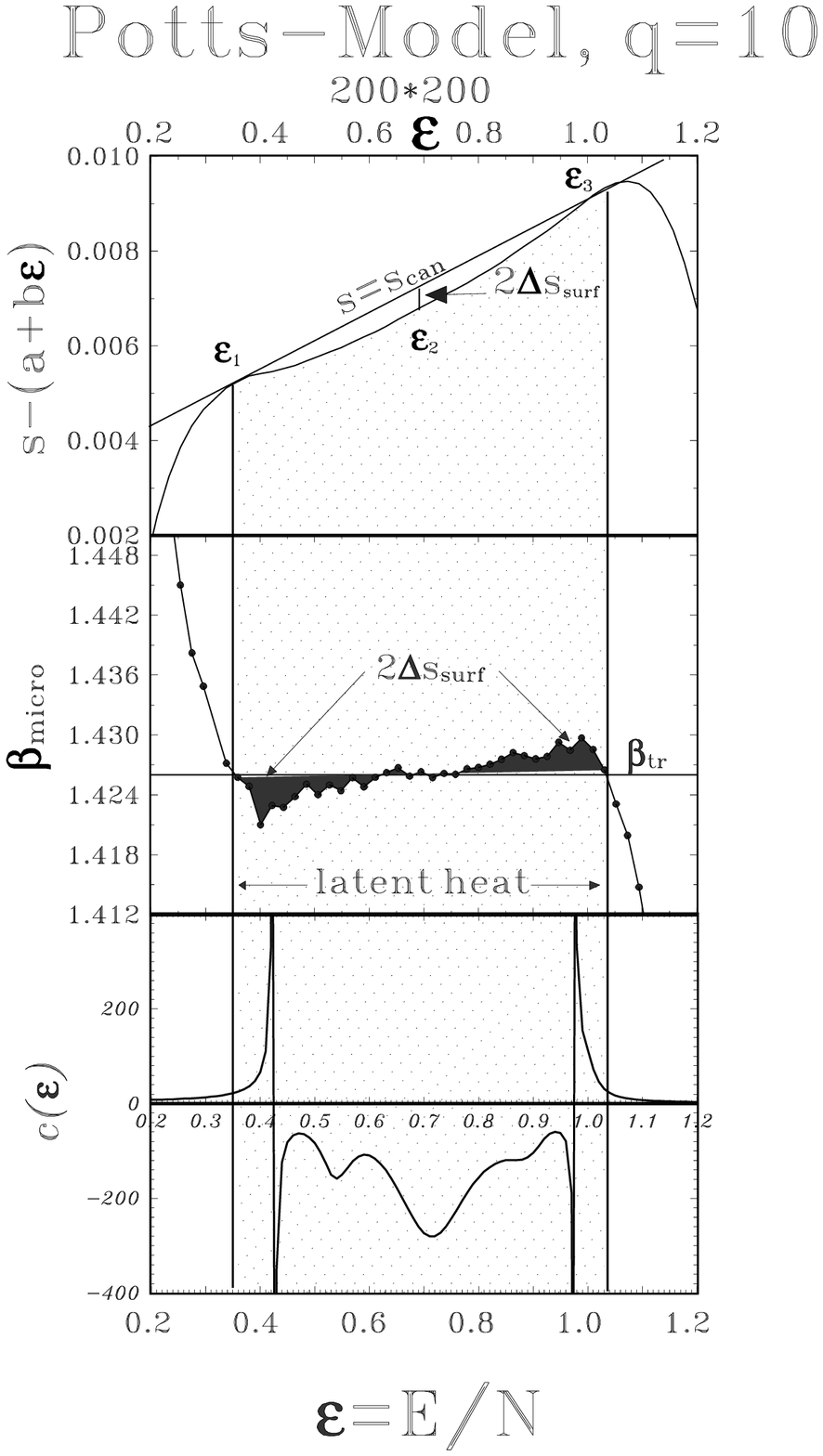}                                         
\end{minipage}\ ~ \
\begin{minipage}[t]{6cm}
{\scriptsize Fig.(2)
 Specific entropy
\protect{$s(\varepsilon)=\int_0^{\varepsilon}{\beta_{micro}(\bar{\varepsilon})
d\bar{\varepsilon}}$} vs.  the specific energy $\varepsilon$ for the 2-dim.
Potts model with $q=10$ on a $200*200$ lattice.  In order to visualize the
anomaly of the entropy the linear function $a+b\varepsilon$ ($a=s(\varepsilon=0.17)$,
$b=1.42$) was subtracted. Because we use periodic boundary conditions one
needs two cuts to separate the phases  and the depth of the convex intruder is
twice the surface-entropy.
\protect\newline
b) Inverse temperature $\beta_{micro}(\varepsilon)=1/T(\varepsilon)$ as directly
calculated by $M\!M\!M\!C$ 
\protect\newline
c) Specific heat $c(\varepsilon)=-\beta^2/(\partial\beta/\partial\varepsilon)$.
The canonical ensemble of the bulk jumps over the shaded region between the
vertical lines at $\varepsilon_1$ and $\varepsilon_3$. This is the region of
the coexistence of two phases one with ordered spins, the other with disordered
spins. Here $c(\varepsilon)$ has two poles and in between it becomes negative.
The canonical thermodynamics is blind to this region. Observe that the poles
are {\em inside} $\varepsilon_1\le\varepsilon\le\varepsilon_3$, i.e. the
canonical specific heat remains finite and positive as it should.
 }
\end{minipage}
\section{Signals of a ``phase transition'' in nuclear fragmentation}

In the review article ref.\cite{gross95} I proposed the caloric equation of
state  $T(\varepsilon)$ for $^{131}$Xe. This had two anomalies
compared to the standard parabolic dependence of a Fermi-gas $T\propto
\sqrt{\varepsilon^*}$, fig.(13).  One is quite pronounced with even a backbending of
$T(\varepsilon^*)$ at $\varepsilon \sim 2.5$MeV, $T\sim 4.5$MeV and a second
one is less pronounced and has no backbending is at $\varepsilon \sim 5$MeV,
$T\sim 6$MeV. The first one had a width $q_{lat}\sim 1$MeV/nucleon and the
second $\sim 1.5$MeV/nucleon and may even be a {\em transition of second
order}(?). The latter one shall not be discussed here. This is likely the
transition found recently at GSI.\cite{pochodzalla95,schwarz96} In the above
review article the first anomaly was compared to the apparent slope temperature
$T_{app}$ of evaporated $\alpha$-particles from $^{32}$S+Ag and of $^{16}$O+Ag
from the Texas A\&M group which show the same narrow anomaly at similar
energies and temperatures.\cite{nebbia86,wada89} However, these data had too
large error-bars to allow any firm conclusion. The interesting aspect of this
first anomaly is that one may see it in theory as well as in experiment in
$\alpha$-{\em evaporation} spectra even though in the model they are linked to
the $\approx$ sudden opening of the IMF production.\\~\\
\noindent
\begin{minipage}[b]{6.75cm}
\includegraphics*[bb = 43 33 478 586, angle=-90, width=6.5cm,  
clip=true]{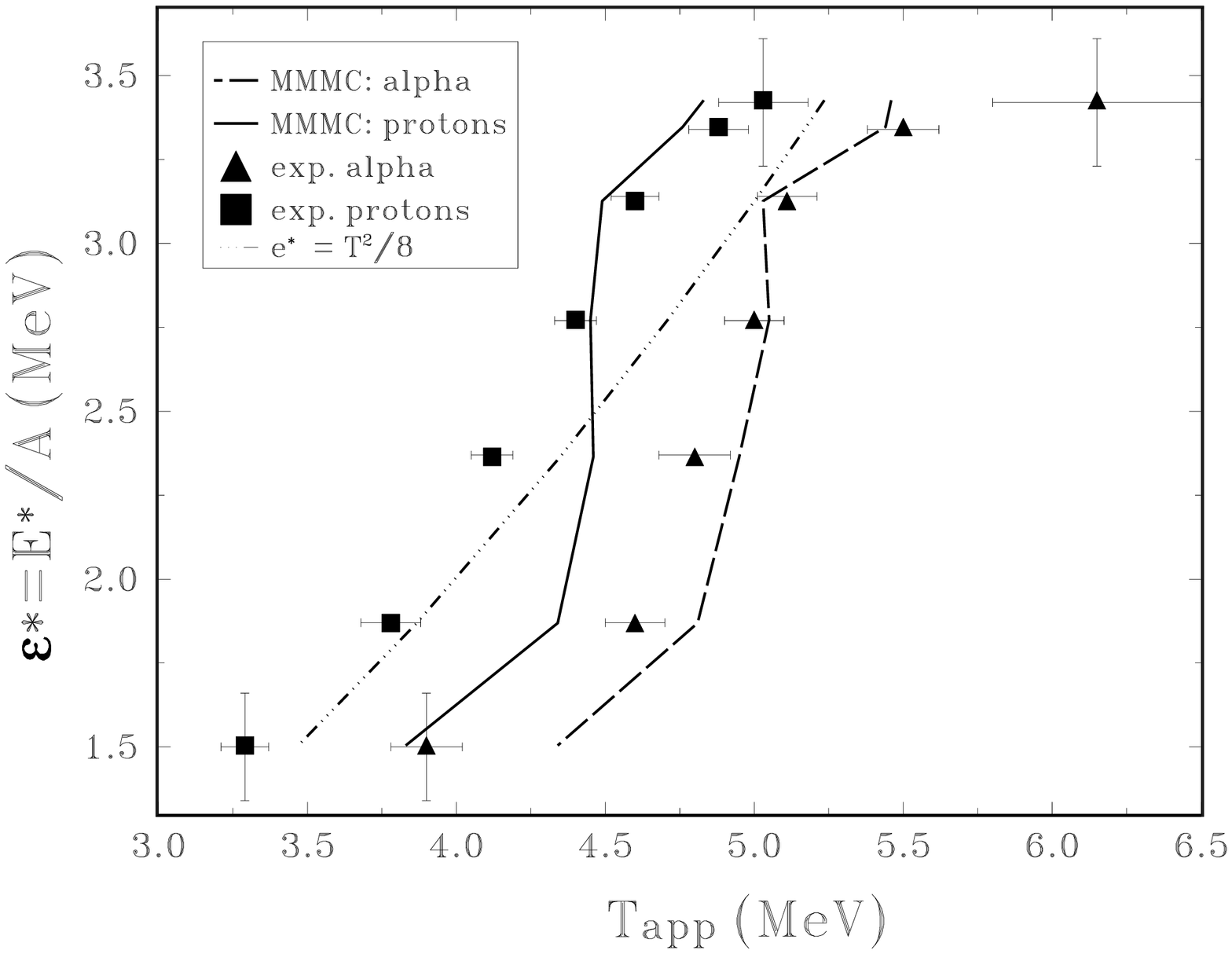}                                         
\\
{\scriptsize Fig.(3)
Experimental and theoretical (with $M\!M\!M\!C$) caloric equation of
state, $T_{app}(\varepsilon^*)$ for $p$ and $\alpha$. The horizontal error-bars
give the statistical uncertainty to extract the slope from the raw spectra in
ref.\protect\cite{chbihi91a}. Different methods to determine the excitation
energy lead essentially to a parallel up or down shift of the curves by the
amount indicated by the vertical bars at the lowest and highest data point. The
dash-dotted curve is a Fermi-gas calculation ($\varepsilon=T^2/8$). 
}
\end{minipage} \ ~ \
\begin{minipage}[b]{4.75cm}
The experiment by Chbihi et al. of incomplete fusion reactions of 701 MeV
$^{28}$Si + $^{100}$Mo gives further evidence of this
transition.\cite{chbihi91a,gross144} We plot in fig.(3) $
\varepsilon^*$, the excitation energy per nucleon vs.  $T_{app}$, where
$T_{app}$ is the slope of the raw evaporation spectra for protons and alpha
particles. The curves give the $T_{app}(\varepsilon^*)$ dependence deduced from
the microcanonical statistical multifragmentation model ($M\!M\!M\!C$)  using
its standard parameters.\cite{gross95,gross144} Also the experimental
uncertainties for the proton and alpha curves are given. The similarity of
the shapes of the experimental and simulated  $T_{app}(\varepsilon^*)$ for the
$\alpha$-spectra is quite evident. The differences between the shapes of these
curves and the parabo\end{minipage}-lic dependence (dotted curve) expected for a simple Fermi
gas is clearly outside the experimental error margins indicating that some
additional degrees of freedom, which are apparently included in the
($M\!M\!M\!C$) model, become significant in this energy range. The proton data
are not so clear. They do indicate a similar anomaly but are more close to the
parabolic Fermi-gas form than to the $M\!M\!M\!C$ curve. Maybe the protons
diffuse too fast out of the expanding soup of fragments and do not explore the
structure of the accessible phase space sensitively enough.

The experimental analysis of the data provides the values of the mass $A$,
charge $Z$, excitation energy $E^*$,  and angular momentum
$L$ of the source.\cite{chbihi91a} The freeze-out radius $R_f$ was taken as
its standard value of $2.2A^{1/3}$~fm, this means that we simulate a ``phase
transition'' at constant volume.  The results of $M\!M\!M\!C$ calculations,
performed by O.Schapiro, 
with these input values, were subjected to the same software filter
as the experimental set-up which, most importantly, selects only those events
with one big residue.  The mass of the residue was chosen to be $A_{res}
\ge 90$, which is close to $A_{res}$ estimated from the experimental data (the
mass of the residue could not be measured).

The theoretical value of $T_{app}$ was extracted from fitting, as was done for
the raw experimental spectra. Similar to the experiment
the individual temperatures for the different particle species are slightly
different in $M\!M\!M\!C$ from the thermodynamic temperature $T_{thd}$ c.f.
figure (3).  This interesting detail is due to the {\em intrinsic
fluctuation of $T_{thd}$} in the microcanonical ensemble --- a result outside
conventional thermodynamics.

The values of $R_f$ and $A_{res}$ do not influence the general shape of the
$T_{app}(\varepsilon^*)$ curves. However, the $T_{app}(\varepsilon^*)$ curves
shift along the 
$T_{app}$ - axis if different values of these parameters are used.  The shifts
produced by reasonable changes in $A_{res}$ are larger than those produced by
reasonable changes in $R_f$. We checked that the anomaly in  $T(\varepsilon)$
is not due to the changes of the angular momentum from $L=18.2$ to $48.8
\hbar$.  It exists also at $L=0$.

While the similarity of the shapes of the experimental and simulated
$T_{app}(\varepsilon^*)$ is quite reasonable for $\alpha$, d, and t-particles,
(see Fig.(3)) significant differences exist in detail.  The simulated curves
for d and t (not shown here) have the {\em same} S-shape as the data
(and the $\alpha$-s) but are
shifted towards lower values of $T_{app}$. The higher $T_{app}$ values of the
experimental deuteron and triton spectra might again be an indication of a
faster diffusion of these less bound fragments from a hotter stage of the early
fragmented system. However, differently to the protons their diffusion is on
the other hand slow enough to realize the transition in the underlying
structure of the phase space. This interpretation is supported by a recent
experiment by ref.\cite{gelderloos95} showing a {\em simultaneous} early
fragmentation and a smaller ``decay'' (diffusion) time for d and t.

The experimental data also suggest an association between the onset of IMF
production and the anomaly in $T_{app}(\varepsilon^*)$. In both theory and
experiment the multiplicity of Li-fragments just starts to rise at the same
excitation.
The success of the $M\!M\!M\!C$ model in showing the same
anomaly in the caloric equation of state in a nearly automatic
manner has a potential far reaching consequence. The central assumption of this
model is an ergodic mixing of the fragmented system filling the accessible
N-fragment phase space with a well defined freeze-out region uniformly.
This is quite different from sequential binary fragmentation, with {\em
independent} motion of fragments outside the binary barrier. Here an anomaly in
$E(T)$ would not have such a natural explanation.

Taking all evidence together: The anomaly in all four spectra (proton,
deuteron, triton, and alpha) at the same excitation energy as predicted by
$M\!M\!M\!C$ and also the earlier data of the Texas A\&M-group, in spite of
their huge error bars there is in my opinion no doubt that this is a signal of
a ``phase transition'' due to the relatively sudden opening of additional phase
space at the onset of fragmentation, the lower one predicted in
ref.\cite{gross95}.  Due to the strong stochastic (dissipative) coupling of the
various fragmentation channels this ``phase transition'' is felt in all channels
even in the pure evaporation channels. The data of the Texas A\&M
group\cite{nebbia86,wada89}, the data of Chbihi et al.\cite{chbihi91a}, as well
as the results of the Aladin collaboration\cite{pochodzalla95} are the first
experimental caloric signals of a phase transition in a nucleus. The data of
ref.\cite{schwarz96} have now smaller error bars and approach the higher
(second order ?) transition mentioned above.\cite{gross95} They show a slow
monotonic {\em increase} of $T$ with rising excitation. I.e. they do not
anymore exhibit the dramatic plateau in $T(E/A)$ with the spectacular
``specific latent heat'' of $\sim 5$ MeV, close to the total binding energy per
nucleon.\cite{schwarz96}

\end{document}